\documentclass[10pt,conference]{IEEEtran}
\IEEEoverridecommandlockouts
\usepackage{cite}
\usepackage{epsfig}
\usepackage{epstopdf}
\usepackage{amsmath,amssymb,amsfonts}
\usepackage{times}
\usepackage{standalone}
\usepackage{graphicx}
\usepackage{textcomp}
\usepackage{xcolor}
\usepackage{algorithm} 
\usepackage{algpseudocode} 
\usepackage{tikz}
\usepackage{caption}
\usepackage{subcaption}
\usetikzlibrary{shapes,arrows}
\usetikzlibrary{arrows.meta,calc,fit,positioning}
\def\BibTeX{{\rm B\kern-.05em{\sc i\kern-.025em b}\kern-.08em
    T\kern-.1667em\lower.7ex\hbox{E}\kern-.125emX}}

\newtheorem{theorem}{Theorem}[section]

\newtheorem{lemma}[theorem]{Lemma}

\begin{document}
\bibliographystyle{IEEEtran}
\title{securePrune:Secure block pruning in UTXO based blockchains using Accumulators\\
\thanks{This research was funded by Indigenous 5G Test Bed (Building an end to end 5G Test Bed) in India, Department of Telecommunication Network $\&$ Technologies (NT) Cell.}
}

\author{\IEEEauthorblockN{ Swaroopa Reddy B}
\IEEEauthorblockA{\textit{Department of Electrical Engineering} \\
\textit{Indian Institute of Technology Hyderabad}\\
Hyderabad, India \\
ee17resch11004@iith.ac.in}
}

\maketitle

\begin{abstract}
In this paper, we propose a scheme called \textit{securePrune} for reducing the storage space of a full node and  synchronization time of bootstrapping nodes joining the Peer-to-Peer (P2P) network in an  \textit{Unspent Transaction Outputs} (\textit{UTXO}) based blockchain like bitcoin using RSA accumulators. The size of the bitcoin blockchain is growing linearly with transactions.  We propose a new block structure to represent the \textit{state} of a blockchain also called \textit{UTXO} set by including an accumulator of a \text{state} in the block header and proofs of knowledge for inclusion and deletion of the transactions of the current block in the block. In our scheme, the miners periodically release a \textit{snapshot} of the blockchain \textit{state}. The other full nodes in the network, securely prune the historical blocks after attaining the required number of confirmations for the \textit{snapshot} block,  which in turn confirms the \textit{snapshot} of the state through an accumulator specified in the block header and proofs inside the block.  The secure and periodic pruning of the old blocks, reduce the synchronization time for a new node joining into the network. The simulation
results demonstrate a significant reduction in the storage space of a full node and bootstrapping cost of the new nodes.
\end{abstract}


\section{Introduction}
The Blockchain is a revolutionary technology behind the  Peer-to-Peer (P2P) cryptocurrency networks like bitcoin \cite{bitcoin} and smart contract enabled P2P networks like Ethereum \cite{ethereum} and Hyperledger \cite{fabric}. The decentralized and trust less nature of the blockchain created a space for applications in healthcare \cite{medical} , Internet of Things (IoT) \cite{IoT}.  

The transactions are the fundamental entities in the blockchain which represents the transfer of coins from one party to another. The miner collects the multiple transactions from other miners/fullnodes and ceates a block with computationally hard problem called Proof-of-Work (PoW) \cite{bitcoin}. The immutability of the blockchain ledger lies in PoW, as any modification to the ledger by a miner needs to recompute all the blocks till the current position of the chain. To achieve the consensus among all the nodes, while creating a new node the miner selects a tip of the longest chain in the network.

The bitcoin blockchain is a P2P network of miners \cite{bitcoin}, fullnodes \cite{fullnode} and simplified payment verifiation nodes (SPV) \cite{bitcoin} . The miners play a key role in generating the blocks through PoW puzzle.
The full node stores all the blocks since the genesis block along with the blockchain state ($UTXO$ set) \cite{UTXO}.
The $UTXO$ set keep track of the all unspent output transactions of the historical blocks and used as sources for new input transactions. A full node contributes to the security of the network through block validation. However, running a full node incurs storage costs as the blockchain data grows exponetially  \cite{blocksize} with time. The main advantage of storing the all blocks by the full node is to make the bootstrapping nodes synchronize with the existing network nodes.


Erasure-code based low storage blockchain node is proposed in \cite{low-storage}. These nodes split every block into fixed size data fragments and generates the coded fragments from the linear combination of the random coefficients generated from the pseudo-random generator. The node can reconstruct the block by the inverse linear combinations. The main limitation of this work is that they
only consider the case when nodes can leave the network or can be unreachable, they do not consider
adversarial nodes that can provide maliciously formed coded fragments. 

A Secure Fountain architecture founded on coding theory is proposed in \cite{SeF} for storage efficiency of the blockchain node by reducing the storage cost and still contribute to bootstrap a new node joining the network. In this scheme, the nodes reduce the storage cost by encoding the validated blocks into a small number of coded blocks using founain codes \cite{LT}. The secure Fountain (SeF) archetecture uses header-chain as a side information to check whether the decoded blocks 
are formed from the mallicious modifications.



In \cite{dynamic}, authors proposed a Dynamic distributed storgae for scaling the blockchains by allocating the nodes into dynamic zones. The nodes in each zone will store a share of private key using shamir's secret sharing \cite{shamir} for encrypting the block data and apply a distributed storage codes such as \cite{exact-regeneration}, \cite{optimal-LRC} for reducing the storage cost. 

However, In all these works, a bootstrapping node need to download all the blocks in the form of distributed coded fragments and validate all the decoded blocks to synchronize with the existing nodes.

%


In this paper, we propose a periodic pruning of the historical blocks based on the security confirmations guaranteed by the RSA accumulator \cite{accumulator}, \cite{batching} of the \textit{UTXO} set and PoW based longest chain consensus algorithm \cite{bitcoin}. 

The main contributions of the paper are as follows.
\begin{enumerate}[]
\item The algorithm for block generation by a miner by adding accumulator for state (\textit{UTXO} set) in the block header to make the state as a part of the PoW consensus algorithm. The miner also includes Non-Interactive proof of Exponentiation (NI-PoE) proofs for inclusion of the new output transations and deletion of the \textit{UTXO} sources of the input transactions of the current block.
\item The algorithm for validation of a block by every full node based on the  NI-PoE proofs \cite{batching} added by the miner inside the block.
\item The periodic pruning of the blocks at regular intervals of the block height based on the security guaranteed by the accumulator state and the NI-PoE proofs.
\item The bootstrapping procedure for synchronizing the new nodes joining the network.
\end{enumerate}

Through the event-driven simulation of blockchain, we have shown the $85\%$ reduction in the storage space of a \textit{securePrune} protocol full node compared to the bitcoin full node and also significant reduction in the synchronization time due to the requirement of validation of less number of historical blocks compared to the validation of all the historical blocks in the bitcoin.

The rest of the paper is organized as follows. In Section II, we present the related work in the reduction of the storage space of a bitcoin full node. Section III gives the system model and notations used in the protocol. Section IV describes the preliminaries for generation of accumulator state and NI-PoE proofs. In Section V, we decribe the proposed protocol for secure and periodic pruning and synchronization of the bootstrapping nodes. In Section VI, we present and discuss the simulation results. Section VII presents the concluding remarks and future works.

\section{Related Work}
The SPV node or light weight node \cite{bitcoin} keep the block headers of a longest PoW chain, instead of the entire blockchain data.  The lightweight node depends on the full node for the verification of a transaction by querying the merkle branch linking the transaction to block where it is time stamped. The block pruning \cite{fullnode} is allowed in bitcoin to store most recent blocks of the chain. However, due to lack of old blocks like in full node, \textit{ pruned node} cannot serve the new full nodes joining the network. The concept of \textit{assumed-valid blocks} \cite{assumed-valid} have been introduced in bitcoin, where a bootstrapping node skips the  script validation of the transactions for ancestors of known-good blocks, without changing the security model. However, the new nodes still need to download entire historical data to create the current state of the blockchain. 

In \textit{coinPrune} \cite{coinprune} protocol, the authors proposed the pruning older blocks by creating a \textit{snapshot} of the state at regular intervals, provided the collective reaffirmations to \textit{snapshot} by the miners. In this protocol, the fullnode prunes the historical blocks provided that the \textit{snapshot} receives the required number of reaffirmations from all the miners. However, there is a possibilty of the \textit{Denial-of-Service} (DOS) attack by the miners in  reaffirming the \textit{snapshot}. So, theere is no gaurantee that pruning will happen at every \textit{reaffirmation window} of a \textit{snapshot} release.

\section{System Model and parameters}
The parameters used in \textit{securePrune} protocol are listed in TABLE \ref{table:symbols}.
\begin{table}[!t]
\renewcommand{\arraystretch}{1.3}
\caption{ \\ Parameters used in the \textit{securePrune} protocol}
\centering
\begin{tabular}{|c|l|}
\hline
\textbf{Symbols}& \textbf{Description} \\
\hline
$hash(.)$ & Cryptographic hash function \\
\hline
$root(.)$ & Merkle root of set of transactions \\
\hline
$H_{prime}(.)$ & Prime representative function \\
\hline
$validate(.)$ & Transaction validation function \\
\hline
$R_i$ & Merkle root of $t_i$ \\
\hline
$B_i$ & $i^{th}$ block of the blockchain \\
\hline
$M$ & Memory pool/set of unconfirmed transactions \\
\hline
$tx$ & Transaction \\
\hline
$t_i$ & $(tx_1, tx_2, ....,tx_{|t_i|})$: Set of transactions in $B_i$ \\
\hline
$S_i$ & State of the block chain ($UTXO$ set) at block $B_i$ \\
\hline
$S_d$ & Set of utxo's spent in the new block \\
\hline
$S_a$ & Set of output transactions in the new block $B_i$ \\
\hline
$A_i$ & Accumulator state \\
\hline
$W$ & List of membership witnesses of \textit{UTXO} set \\
\hline
$\pi_d$ & NI-PoE proof for deletion of set $S_d$ from accumulator \\
\hline
$\pi_a$ &  NI-PoE proof for addition of set $S_a$ to accumulator \\
\hline
$T_{proofs}$ & NI-PoE verification time \\
\hline
$\Delta_s$ & Number of blocks between two \textit{snapshots} \\
\hline
$n$ & Number of nodes in the network \\
\hline
$q$ & Fraction of the attacker's hashrate \\
\hline
$n_p$ & Number of peers connected to each node \\
\hline
$T_p$ & Propagation delay \\
\hline
$b$ & Block size in \textbf{MB} \\
\hline
$R$ & Average download bandwidth \\
\hline
$R_v$ & Average validation rate of a block \\
   \hline
$p_v$   & Computational power with node $v$\\
\hline
$\lambda$ & Block creation rate \\
\hline
$D$ & End-to-end delay in the network \\
\hline
$k$ & Number of confirmations \\
\hline
$h$ & Height of the blockchain\\
\hline
$m$ & Number of mining nodes in the network \\
\hline
\end{tabular}
\label{table:symbols}
\end{table}

%

\subsection{Overview of the transactions and UTXO set}

There are two types of the transactions in every transaction of a block - inputs and outputs \cite{bitcoin}. The inputs specifies the previous transaction outputs as sources of the bitcoins in a transaction and the outputs are the destination of bitcoin transfer. Each transaction contains multiple inputs and multiple outputs to combine and split the values of the coin transfer.  The full node stores the $UTXO$ set in the chainstate database of the Bitcoin core \cite{bitcoincore}. The database consists of records of key-value pairs \cite{UTXO} . The key of the record is transaction hash and the value stores the transaction information.  Every record in the \textit{UTXO} set represent the outputs yet to be spent in future transactions.

Let at a block height $i$, every full node in the blockchain stores a copy of the state (\textit{UTXO} set) $S_i$ repersented as
\begin{equation}
S_i = \{u_j:j = 1,2,\dots ,|S_i|\}
\end{equation}
Where, $u_j$ is a record in the \textit{UTXO} set 

However, for every new block ($B_i$) addtion to the chain, the state of the full node changes with the transaction set $t_i$ as described in state transition Algorithm in Section \ref{state_transition}.

\subsection{The modified block structure in the proposed protocol}
The blockchain at a height $h$ is modeled as a vector of blocks represented as 
\begin{equation}
C_h = (B_0,B_1,\dots , B_h)
\end{equation}
where, each block $B_i$ is a tuple consists of block header ($H_i$), NI-PoE proof ($\pi_d$) for deletion of set $S_d$ from the \textit{UTXO} set and NI-PoE proof ($\pi_a$) for addition of set $S_a$ to the $UTXO$ set. While generating a new block, the miner includes a list of transactions $t_i$ into the block from the transaction memory pool ($M$) stored with every miner and full nodes.
\begin{equation}
B_i = \hspace{0.2cm} <H_i, (A'_{i-1}, \pi_d, \pi_a), t_i>
\end{equation}
Where, the tuple $(A'_{i-1}, \pi_d, \pi_a)$ results from the state transition of the \textit{UTXO} set.

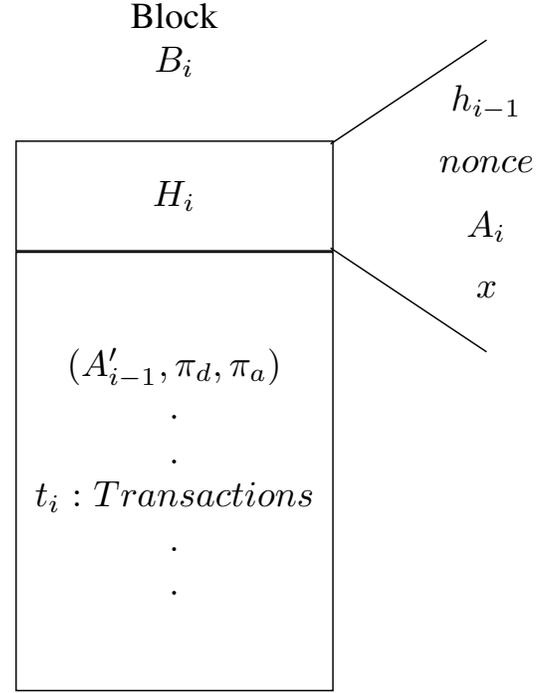
\begin{figure}[t]
\centering
\resizebox{\columnwidth}{!}{
\tikzstyle{block} = [square, draw,
    text width=7em, text centered, minimum height=4em]
\tikzstyle{sum} = [draw, circle, node distance=3cm]
\tikzstyle{input} = [coordinate]
\tikzstyle{output} = [coordinate]
\tikzstyle{pinstyle} = [pin edge={to-,thin,black}]
\tikzstyle{line} = [draw, -latex']

\tikzset{
   bigbigbox/.style = {minimum width=2.5cm, rectangle},
   bigbox/.style = {draw,rectangle},
   box/.style = {minimum width=2.7cm, rounded corners,rectangle, fill=blue!20},
   square/.style = {minimum width=15mm,minimum height=15mm}
   }

\begin{tikzpicture}[auto, node distance=2cm,>=latex']
\node [rectangle, draw,text width=8em, text centered, minimum height=3em,name=genesis] {$H_i$} ;

\node [rectangle, draw,text width=8em, text centered, minimum height=12em, below of=genesis,node distance=2.65cm] (1) {$(A'_{i-1}, \pi_d, \pi_a)$ \\$.$\\$.$ \\ $t_i:Transactions$  \\ $.$ \\ $.$};

\node [ text width=3em, text centered, minimum height=3em,  above of=genesis,node distance=1.5cm] () {Block $B_i$};
%
\draw (1.5,0.5) -- (3.0,1.5) node[](a) {};
\draw (1.5,-0.5) -- (3.0,-1.5) node[](b) {};
\node [text width=3em, text centered, minimum height=1em, below of=a,node distance=0.6cm] (2) {$h_{i-1}$};
\node [text width=3em, text centered, minimum height=1em, below of=2,node distance=0.6cm] (3) {$nonce$};

\node [text width=3em, text centered, minimum height=1em, below of=3,node distance=0.6cm] (4) {$A_i$};

\node [text width=3em, text centered, minimum height=1em, below of=4,node distance=0.6cm] (5) {$x$};

\end{tikzpicture}
}
\caption{\textit{securePrune} Block structure}
\label{fig:block_diagram}
\end{figure}

In addition to the elements of the bitcoin block header, our proposed model includes an extra element called the accumulator state ($A_i$), which is an RSA accumulator \cite{accumulator} to represent the sate of the blockchain ($S_i$) in the block header.
\begin{equation}
H_i = (h_{i-1}, nonce, A_i, x)
\end{equation}
where, $h_{i-1} = hash(H_{i-1})$, $nonce$ is a variable to solve the PoW puzzle and $x$ is the other meta data (like version, time, difficulty etc) similar to bitcoin block header \cite{bitcoin}. The modified structure of the block is shown in Fig. \ref{fig:block_diagram}.

\section{Preliminaries}
The following definitions of RSA accumulators \cite{accumulator}, \cite{batching} are used in our
work.

\textbf{Definition 1. (Accumulator of State).}
Let $\mathbb{G}$ be a group of unknown order  and $g \in \mathbb{G}$, the accumulator state of a block $B_i$ is an RSA accumulator \cite{accumulator} of the unspent transaction outputs present in $UTXO$ set $S_i = \{u_j:j = 1,2,...,|S_i|\}$ and is computed as 
\begin{equation}
A_i = \prod^{|S_i|}_{j=1} g^{U_j}
\end{equation}
Where, $U_j$ is the prime representative of the element $u_j$ \cite{batching}
\begin{equation}
U_j = H_{prime}(u_j)
\end{equation}
The dynamic accumulator \cite{dynamic_acc} is an accumulator that allows to add or delete elements to the accumulator.

\textbf{Definition 2. (Membership Witness).}
A membership witness is simply an accumulator without an aggregated element. The membership witness for $u_m$ is defined as
\begin{equation}
W_m = \prod^{|S_i|}_{j=1,j \neq m} g^{U_j}
\label{eq:w}
\end{equation}
\textbf{Definition 3. (Shamir Trick).}
Let $x,y \in S$ and $g \in \mathbb{G}$, the membership witnesses for $x$ and $y$ are the $x^{th}$ root of $g$ and $y^{th}$ root of $g$, then the \textit{ShamirTrick} \cite{shamirtrick}, \cite{batching} is a $(xy)^{th}$ root of the group element $g$ from the Bezout's coefficients of $x$ and $y$.

While creating a new block, the miner generates new \textit{accumulator state} $A_i$ from $A_{i-1}$ in two stages - The deletion of the set $S_d$ from the accumulator $A_{i-1}$ followed by addition of set $S_a$ to obtain the new state $A_i$.
\begin{equation}
A_i = BatchAdd(BatchDel(A_{i-1},S_d),S_a)
\end{equation}
\textbf{Definition 4. (Batch Deletion (BatchDel)).}
Let $S_d$ represent the set of sources for inputs of the transactions in the new block $B_i$, the state $S_{i-1}$ needs to delete the records $S_d$ from the database.
The deletion of the set $S_d$ from accumulator state $A_{i-1}$ can be obtained from $BatchDel$ \cite{batching}. The $BatchDel$ uses the membership aggregation function $AggMemWit$   \cite{batching} to compute the aggregate membership witness of all elements in $S_d$ from the individual membership witnesses of each element. 
The $AggMemWit$  is simply an accumulator without  elements of set $S_d$.
\begin{equation}
A'_{i-1} = W_{agg} = \prod_{u_j \in S_{i-1}\backslash S_d} g^{U_j}
\label{eq:w_agg}
\end{equation}
where, $W_{agg}$ is the aggregated membership witness of all the elements of the set $S_d$ generated by Shamir Trick \cite{batching}. 
The $BatchDel$ gives the intermediate state of the accumulator $A'_{i-1}$ to process further for obtaining the new \textit{accumulator state} ($A_i$) of the new block $B_i$ from the set $S_a$.

\textbf{Definition 5. (Batch Addition (BatchAdd).}
The addition of the elements of set $S_a$ to accumulator state requires a batch addition $BatchAdd$ \cite{batching} for efficient computation.
\begin{equation}
A_i = (A'_{i-1})^{U^*}
\end{equation} 
where,
\begin{equation}
U^* = \prod_{s \in S_a} H_{prime}(s)
\end{equation}
\textbf{Definition 6. (Proof of Exponetiation (PoE) \cite{batching}).}
Let $\mathbb{G}$ be a group of unknown order and $u, w \in \mathbb{G}$, the proof of exponetiation in the Group $\mathbb{G}$, when both the prover and verifier are given $(u,w, x \in \mathbb{Z})$ and the prover wants to convince the verifier that $w=u^x$.

The Non-interactive PoE (NI-PoE \cite{batching}) proofs $\pi_d$ and $\pi_a$ are generated during the batch updates for the efficient verification without any interaction between prover(miner) and verifier(full node).
\begin{equation}
\pi_d = NI-PoE(u,x,w)
\end{equation}
\textbf{Definition 7. (Updating membership witnesses).}
The intermediate accumulator state \eqref{eq:w_agg} denotes the membership witness for all the elemets of the set $S_d$. Let $s \in S_{i-1} \backslash S_d$ and $w_s$ is the membership witness of $s$ before deletion of set $S_d$ as per \eqref{eq:w}, then the updated membership witnesses for all $s \in S_{i-1} \backslash S_d$ are generated as follows
\begin{equation}
w'_s = ShamirTrick(A'_{i-1}, w_s, \prod_{x \in S_d}x, s)
\end{equation}
The memebrship witness updates for all $s \in S_{i-1} \backslash S_d$ after the addition of elements of the set $S_a$ are calculated as follows
\begin{equation}
w''_s = (w'_s)^{\prod_{x \in S_a}x}
\end{equation}
The membership witnesses for  elements $x \in S_a$ are calculated as follows
\begin{equation}
w_x = (A'_{i-1})^{\prod_{y \in S_a, y \neq x}y}
\end{equation}
\section{Secure block pruning protocol}
In this section, we discuss the proposed secure pruning protocol for storage scalability and the synchronization process of the bootstrapping nodes. The protocol requires the modification in the block generation procedure by the miners and the validation procedure of a block by the full nodes in the network based on the  accumulators and NI-PoE.
\subsection{{Requirements of the securePrune protocol}}
\label{requirements}
\subsubsection{State transition Algorithm}
\label{state_transition}
\begin{algorithm}[h]
  \caption{State transition Algorithm}\label{Algorithm1}
  \textbf{Input:} $S_{i-1}$, $t_i$  \\
	\textbf{output:} $S'$ - new state, $S_d$, $S_a$
  \begin{algorithmic}[1]
    \Procedure{stateTransistion }{$S_{i-1}$, $t_i$}
      \State $S' \gets S_{i-1}$
      \For{$tx$ in $t_i$}
      		\State $isValid \gets validate(tx)$
      		\If{$isValid$}
      		\For{\textit{input} in \textit{tx}}
             	\State $id \gets input[txHash]$
             	\State delete $u_{j}[id]$ from  $S'$
             	\State $S_d.append(u_{j}[id])$
             	
             \EndFor
             \For{$output$ in $tx$}
             	\State $S'$.append($output$)
             	\State $S_a$.append($output$)
             \EndFor
             \Else
             	\State return \textit{False}
             	
             \EndIf
	  \EndFor
	   \State \textbf{return} $S'$, $S_d$, $S_a$ 
    \EndProcedure
  \end{algorithmic}
\end{algorithm}

The $UTXO$ set (state) of the blockchain is dynamic and changed for every new block addition to the blockchain. The following algorithm describes the transition of a miner (or full node) while generating a new block (or after receiving a new block). The new state transition function returns the set of deleted elements ($S_d$) and added elements ($S_a$) along with the new $UTXO$ set. 

\subsubsection{Modified PoW Algorithm}

The modified Proof-of-Work function for mining a new block is described in Algorithm \ref{Algorithm2}. This PoW funtion includes Accumulator state $A_i$ along with other parameters into the block header for providng immutable blockchain state $S_i$. It also includes NI-PoE proofs ($\pi_d$, $\pi_a$) for \textit{deletion} and \textit{addition} of the new set of elements ($S_d$, $S_a$) to the state from the present transaction set $t_i$.

The NI-PoE proof $\pi_d$ is obtained from the $BatchDel$ function \cite{batching} as a proof for deletion of the unspent transactions refered in the \textit{inputs} of the set $t_i$. The $BatchDel$ function deletes the sources of \textit{inputs} ($S_d$) from the accumulator state of the previous block $A_{i-1}$ and generates the NI-PoE proof $\pi_d$.
The proof $\pi_a$ is also an NI-PoE proof generated from the $BatchAdd$ function \cite{batching} for adding the \textit{outputs} ($S_a$) of set $t_i$ .   
\begin{algorithm}[h]
  \caption{The modified \textit{PoW} function \\ for the secure prune protocol}\label{Algorithm2}
  \textbf{Input:} $S_{i-1}$, $C_{i-1}$, $M$,$W$  \\
	\textbf{output:} $C_i$
  \begin{algorithmic}[1]
    \Procedure{securePrunePoW}{$S_{i-1}$, $C_{i-1}$}
      \For{$tx$ in $M$}
      	  \State $t_i$.append($tx$)
      	  \If {size of $B_i$ $>$ Max Block Size}
      	  		\State break
      	  \EndIf
      \EndFor
       
      \State $S_i, S_d, S_a \gets  	
              stateTransation(S_{i-1},t_i)$
       \State $A'_{i-1}, \pi_d \gets BatchDel(A_{i-1}, S_d,W)$
       \State $A_i, \pi_a \gets  BatchAdd(A'_{i-1},S_a)$
       \State $W' = updateMemWit(A'_{i-1},W,S_d,S_a)$
	   \State $nonce \gets 0$
	   \While{$nonce < 2^{32}$}
			\State $h \gets hash(H_{i-1},R_{t_i},A_i,x)$
			\If{$hash(nonce,h) > Difficulty$}
				\State break
			\EndIf
			\State $nonce \gets nonce + 1$
		\EndWhile
		\State $W \gets W'$
		\State $H_i \gets \hspace{0.1cm}
		 <H_{i-1},R_{t_i},A_i,x,nonce>$
		 
		\State $B_i \gets \hspace{0.1cm} <H_i,\pi_d,\pi_a,t_i>$
		\State $C_i \gets C_{i-1}B_i $	   
	   \State \textbf{return} $C_i$
    \EndProcedure
  \end{algorithmic}
\end{algorithm}

\subsubsection{Block Validation Algorithm}
We defined a validation function in Algorithm \ref{Algorithm3} to check the validity of $A_i$, $t_i$, $R_i$, $\pi_d$ and $\pi_a$ from the present state $S_{i-1}$, local chain $C_{i-1}$ and the received new block ($B_i$). 
If $B_i$ is valid, the full node adds $B_i$ to $C_{i-1}$, otherwise discards the block. 
\begin{algorithm}[h]
  \caption{Block Validation Algorithm}\label{Algorithm3}
  \textbf{Input:} $S_{i-1}$, $C_{i-1}$, $B_i$  \\
	\textbf{output:}  $C_i$, $S_i$
  \begin{algorithmic}[1]
    \Procedure{validateBlock}{$S_{i-1}$, $C_{i-1}$, $B_i$}
      \State $t_i \gets B_i[t_i]$
      \State $count \gets 0$
      \For{$tx$ in $t_i$}
      		\State $isValid \gets validate(tx)$
             	\If{not $isValid$}
             		\State \textbf{return} \textit{False}
             	\EndIf
             \State	$count \gets count+1$
        \EndFor
        \If{$R_i \neq root(t_i)$}
        	\State \textbf{return} \textit{False}
        \EndIf
        \If{$count == |t_i|$}
        \State $A'_{i-1}, \pi_d, \pi_a  \gets B_i$
        \State $A_{i-1} \gets B_{i-1}[accState]$
        \State $S_i, S_d, S_a \gets  	
              stateTransation(S_{i-1},t_i)$
        \State $a \gets NI-PoE.Verify(\prod_{s \in S_d}s, 	               A'_{i-1}, A_{i-1}, \pi_d)$
        \State $b \gets NI-PoE.Verify(\prod_{s \in S_a}s, 	               A'_{i-1}, A_{i}, \pi_a)$
        \EndIf
        \If{$a \wedge b$}
        \State $S_i \gets S'$
        \State $C_i \gets C_{i-1}B_i$
        \EndIf
        \State \textbf{return} $C_i$
    \EndProcedure
  \end{algorithmic}
\end{algorithm}

\subsection{securePrune Protocol}
The protocol differs from the bitcoin protocol by issuing a \textit{snapshot} of the $UTXO$ set at regular intervals of every $\Delta_s$ blocks called \textit{snapshot interval}. The miners while creating a new block as per the Algorithm \ref{Algorithm2} at a height $c\Delta_s$ ($c = 1,2,3,\dots$) releases the \textit{snapshot} along with the block $B_p$ created at that particular height. The snapshot conststs of  an \textit{indetifier} and a copy of the \textit{state} ($S_p$) consists of all the unspent tranasctions including the unspent transactions of the current block. The \textit{snapshot identifier} is the accumulator state present in the block header of \textit{snapshot} block $B_p$.  The chain subsequent to the snapshot block $B_p$ is termed as the \textit{tailchain}. The full node follows the Algorithm \ref{Algorithm3} for validation of a block created during $\Delta_s$  (present in the \textit{tailchain}) by verifying the NI-PoE proofs $\pi_d$ and $\pi_a$, \textit{merkle root} $R_i$ and transactions $t_i$. 

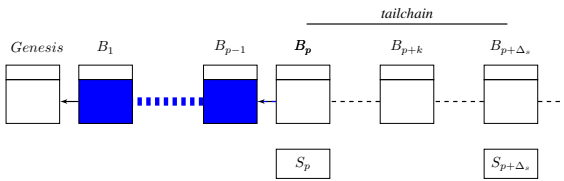
\begin{figure}[h]
\centering
\resizebox{\columnwidth}{!}{
\tikzstyle{block} = [square, draw,
    text width=7em, text centered, minimum height=4em]
\tikzstyle{sum} = [draw, circle, node distance=3cm]
\tikzstyle{input} = [coordinate]
\tikzstyle{output} = [coordinate]
\tikzstyle{pinstyle} = [pin edge={to-,thin,black}]
\tikzstyle{line} = [draw, -latex']

\tikzset{
   bigbigbox/.style = {minimum width=2.5cm, rectangle},
   bigbox/.style = {draw,rectangle},
   box/.style = {minimum width=2.7cm, rounded corners,rectangle, fill=blue!20},
   square/.style = {minimum width=15mm,minimum height=15mm}
   }

\begin{tikzpicture}[auto, node distance=2cm,>=latex']
\node [rectangle, draw,text width=3em, text centered, minimum height=1em,name=genesis] {} ;

\node [rectangle, draw,text width=3em, text centered, minimum height=3em, below of=genesis,node distance=0.7cm] (1) {};
  
\node [rectangle, draw,text width=3em, text centered, minimum height=1em, right of=genesis,node distance=1.75cm] (2) {};

\node [rectangle, draw,fill = blue, text width=3em, text centered, minimum height=3em, right of=1,node distance=1.75cm] (3) {};
  
\draw [->,thick] (3) -- node {} (1);

\node [rectangle, draw,text width=3em, text centered, minimum height=1em, right of=2,node distance=3.0cm] (4) {};

\node [rectangle, draw,fill = blue,text width=3em, text centered, minimum height=3em, right of=3,node distance=3.0cm] (5) {};
  
\draw [dashed,line width = 2mm,color=blue] (5) -- node {} (3);

\node [text width=3em, text centered, minimum height=3em, above of=1,node distance=1.3cm] (A) {$Genesis$};
\node [text width=3em, text centered, minimum height=3em, above of=3,node distance=1.3cm] (B1) {$B_1$};
\node [text width=3em, text centered, minimum height=3em, above of=5,node distance=1.3cm] (B2) {$B_{p-1}$};

\node [rectangle, draw,text width=3em, text centered, minimum height=1em, right of=4,node distance=1.75cm] (6) {};

\node [rectangle, draw,text width=3em, text centered, minimum height=3em, right of=5,node distance=1.75cm] (7) {};

\draw [->,thick] (7) -- node {} (5);
\node [text width=3em, text centered, minimum height=3em, above of=7,node distance=1.3cm] (B3) {$B_p$};

\node [rectangle, draw,text width=3em, text centered, minimum height=1em, right of=6,node distance=2.5cm] (8) {};

\node [rectangle, draw,text width=3em, text centered, minimum height=3em, right of=7,node distance=2.5cm] (9) {};
\node [rectangle, draw,text width=3em, text centered, minimum height=2em, below of=7,node distance=1.5cm] (snap1) {$S_p$};

\draw [dashed,thick] (9) -- node {} (7);
\node [text width=3em, text centered, minimum height=3em, above of=9,node distance=1.3cm] (B4) {$B_{p+k}$};

\node [rectangle, draw,text width=3em, text centered, minimum height=1em, right of=8,node distance=2.5cm] (10) {};

\node [rectangle, draw,text width=3em, text centered, minimum height=3em, right of=9,node distance=2.5cm] (11) {};

\draw [dashed,thick] (11) -- node {} (9);
\node [text width=3em, text centered, minimum height=3em, above of=11,node distance=1.3cm] (B5) {$B_{p+\Delta_s}$};
\node [text width=3em, text centered, minimum height=3em, right of=11,node distance=2.0cm] (x) {};
\draw [dashed,thick] (11) -- node {} (x);

\node [rectangle, draw,text width=3em, text centered, minimum height=2em, below of=11,node distance=1.5cm] (snap2) {$S_{p+\Delta_s}$};

\draw [dashed,thick,color=blue] (7) -- node {} (5);
\node [text width=3em, text centered, minimum height=3em, above of=7,node distance=1.3cm] (B3) {$B_p$};

\node [text width=3em, text centered, minimum height=3em, above left of=B3,color=blue,node distance=0.8cm] (B6) {};
\node [text width=3em, text centered, minimum height=3em, above right of=B5,node distance=0.8cm] (B7) {};
\draw [-,thick] (B6) -- node {\textit{tailchain}} (B7);

\node [text width=3em, text centered, minimum height=3em, below of=B6,node distance=0.5cm] (B8) {};
\node [text width=3em, text centered, minimum height=3em, below of=B7,node distance=0.5cm] (B9) {};

\end{tikzpicture}
}
\caption{Overview of \textit{securePrune} protocol: The blue colour blocks are pruned after attaining a $k$ confirmations to block $B_p$.}
\label{fig:block}
\end{figure}

The full nodes in the network prune all the historical blocks prior to the snapshot block $B_p$, provided that the block $B_p$  achieved $k$ number of confirmations from the \textit{tailchain} blocks created in the network. The full nodes choose the tip of the longest chain similar to bitcoin \cite{bitcoin} for deciding the number of confirmations on $B_p$. Suppose, more than one miner releases a \textit{snapshot} at a height $p+c\Delta_s$, the \textit{snapshot} with longest \textit{tailchain} is a valid \textit{snapshot}. Fig. \ref{fig:block} describes the overview of the \textit{secureprune} protocol.

\begin{lemma}
Let $k$ be the number of confirmations required for a block with very low probability of double-spend to succeed by an attacker, then the number of confirmations required for a \textit{snapshot} is also $k$.
\label{lemma1}
\end{lemma}
\begin{IEEEproof}
If any attacker tries to modifies a transaction in a block, the \textit{hash} of the block change as the  \textit{merkle root} is a function of all the transactions in a block \cite{bitcoin}. The double-spend attack is  the creation of a secret chain longer than the chain with the original transaction. So, an attacker needs to create a chain longer than the honest chain to modify any transaction of a particular block. The probability of double-spend to be succeed by an attacker with a fraction of hashrate $q$ and for a given number of confirmations  is shown in \cite{bitcoin} and \cite{Rosenfeld}. 

Suppose, a miner in the network creates a block at height $p$ and has recieved $k$ number of confirmations. Let $B_{p+1}, B_{p+2}, \dots, B_{p+k}$ are the blocks that confirms block $B_p$. A transaction in the block is said to be a part of the valid chain, if it has $k$ number of confirmations with very low probability of double-spend by an attacker.
 
Let $\bullet$ denotes the state transition function described in Alogorithm \ref{Algorithm1}, the new state of a \textit{snapshot} $S_p$ after $k$ number of blocks is appended to block $B_p$ is given by

\begin{align*}
S_{p+k} &= S_{p+k-1} \bullet B_{p+k} \\
&= S_{p+k-2} \bullet B_{p+k-1} \bullet B_{p+k} \\
&= S_p \bullet B_{p+1} \bullet B_{p+2} \bullet \dots \bullet B_{p+k}
\end{align*}
since the state is represented as a part of the PoW function in Algorithm \ref{Algorithm2} in terms of the accumulator state $A_i$ for a block $B_i$, the \textit{hash} of the block changes with change in the state particularly \textit{snapshot}.

Thus, the number of confirmations required for immutability of the \textit{snapshot} is same as the number of confirmations required for a block against the double-spend attack.
\end{IEEEproof}  
\subsection{Size of the Blockchain in \textit{securePrune}}
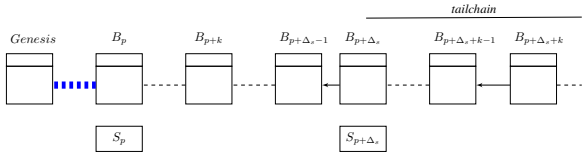
\begin{figure}[t]
\centering
\resizebox{\columnwidth}{!}{
\tikzstyle{block} = [square, draw,
    text width=7em, text centered, minimum height=4em]
\tikzstyle{sum} = [draw, circle, node distance=3cm]
\tikzstyle{input} = [coordinate]
\tikzstyle{output} = [coordinate]
\tikzstyle{pinstyle} = [pin edge={to-,thin,black}]
\tikzstyle{line} = [draw, -latex']

\tikzset{
   bigbigbox/.style = {minimum width=2.5cm, rectangle},
   bigbox/.style = {draw,rectangle},
   box/.style = {minimum width=2.7cm, rounded corners,rectangle, fill=blue!20},
   square/.style = {minimum width=15mm,minimum height=15mm}
   }

\begin{tikzpicture}[auto, node distance=2cm,>=latex']
\node [rectangle, draw,text width=3em, text centered, minimum height=1em,name=genesis] {} ;

\node [rectangle, draw,text width=3em, text centered, minimum height=3em, below of=genesis,node distance=0.7cm] (1) {};
  
\node [rectangle, draw,text width=3em, text centered, minimum height=1em, right of=genesis,node distance=2.5cm] (2) {};

\node [rectangle, draw, text width=3em, text centered, minimum height=3em, right of=1,node distance=2.5cm] (3) {};
  
\draw [dashed,line width = 2mm,color=blue] (3) -- node {} (1);
%
%
%
%
%
\node [text width=3em, text centered, minimum height=3em, above of=1,node distance=1.3cm] (A) {$Genesis$};
\node [text width=3em, text centered, minimum height=3em, above of=3,node distance=1.3cm] (B1) {$B_p$};

%
\node [rectangle, draw,text width=3em, text centered, minimum height=1em, right of=2,node distance=2.5cm] (4) {};
\node [rectangle, draw,text width=3em, text centered, minimum height=3em, right of=3,node distance=2.5cm] (5) {};
\draw [dashed,thick] (5) -- node {} (3);
\node [text width=3em, text centered, minimum height=3em, above of=5,node distance=1.3cm] (B3) {$B_{p+k}$};

\node [rectangle, draw,text width=3em, text centered, minimum height=1em, right of=4,node distance=2.5cm] (a) {};
\node [rectangle, draw,text width=3em, text centered, minimum height=3em, right of=5,node distance=2.5cm] (b) {};
\draw [dashed,thick] (b) -- node {} (5);
\node [text width=3em, text centered, minimum height=3em, above of=b,node distance=1.3cm] (B) {$B_{p+\Delta_s-1}$};

\node [rectangle, draw,text width=3em, text centered, minimum height=1em, right of=a,node distance=1.8cm] (6) {};
\node [rectangle, draw,text width=3em, text centered, minimum height=3em, right of=b,node distance=1.8cm] (7) {};
\node [rectangle, draw,text width=3em, text centered, minimum height=2em, below of=3,node distance=1.5cm] (snap1) {$S_p$};
\node [rectangle, draw,text width=3em, text centered, minimum height=2em, below of=7,node distance=1.5cm] (snap2) {$S_{p+\Delta_s}$};

\draw [->,thick] (7) -- node {} (b);
\node [text width=3em, text centered, minimum height=3em, above of=7,node distance=1.3cm] (B4) {$B_{p+\Delta_s}$};

\node [rectangle, draw,text width=3em, text centered, minimum height=1em, right of=6,node distance=2.5cm] (8) {};
\node [rectangle, draw,text width=3em, text centered, minimum height=3em, right of=7,node distance=2.5cm] (9) {};
\draw [dashed,thick] (9) -- node {} (7);
\node [text width=3em, text centered, minimum height=3em, above of=9,node distance=1.3cm] (B5) {$B_{p+\Delta_s+k-1}$};

\node [rectangle, draw,text width=3em, text centered, minimum height=1em, right of=8,node distance=2.25cm] (10) {};
\node [rectangle, draw,text width=3em, text centered, minimum height=3em, right of=9,node distance=2.25cm] (11) {};
\draw [->,thick] (11) -- node {} (9);
\node [text width=3em, text centered, minimum height=3em, above of=11,node distance=1.3cm] (B6) {$B_{p+\Delta_s+k}$};

%
%
\node [text width=3em, text centered, minimum height=3em, right of=11,node distance=2.0cm] (12) {};
\draw [dashed,thick] (11) -- node {} (12);

\node [text width=3em, text centered, minimum height=3em, above left of=B4,node distance=0.8cm] (B7) {};
\node [text width=3em, text centered, minimum height=3em, above of=12,node distance=1.9cm] (B8) {};
\draw [-,thick] (B7) -- node {\textit{tailchain}} (B8);

\end{tikzpicture}
}
\caption{The blue colured dashed line shows the pruned blocks in the chain}
\label{fig:blocksize}
\end{figure}

Fig. \ref{fig:blocksize} shows the total blocks from block $B_p$ to the current height including $B_{p+\Delta_s}$, block at snap shot $S_{p+\Delta_s}$. Suppose, the nodes pruned the blocks till height $B_{p-1}$ after acheiving the required number of confirmations to \textit{snapshot} $S_p$, then every node in the network stores the blocks from $B_p$ onwards till the current height. 

Suppose, a miner broadcast a block into the network at $p+\Delta_s$ along with \textit{snapshot} $S_{p+\Delta_s}$, then the total number of blocks till block height $p+\Delta_s+k-1$ are $\Delta_s + k$. At height $p+\Delta_s+k$, the nodes prune the blocks $B_p$ to $B_{p+\Delta_s-1}$.
So, the total number of blocks stored with a node is upper bounded by $\Delta_s + k$.

\subsection{Synchronization of the Bootstrapping nodes}
The new node joining the network bootstrap in three steps -
First, it obtains the most recent \textit{snapshot} with the longest \textit{tailchain}. Second, the new node downloads the entire \textit{headerchain} since the \textit{genesis} block and verifies the validity of the \textit{headerchain}. Third, the node downloads the \textit{tailchain} from its peers and validate all the blocks since the \textit{snapshot} block to obtain its  \textit{state}. 

Let $S_p$ is the most recent \textit{snapshot} and a node joins the network at height $h$, then the state of the new node at height $h$ is obtained as
\begin{equation}
S_h = S_p \bullet B_{p+1} \bullet \dots \bullet B_h
\end{equation}

Let $b$, $c$, $s$ are the new nodes joining in \textit{bitcoin}, \textit{coinPrune} and \textit{securePrune} respectively, then, $n_b$, $n_c$, $n_s$ are the number of blocks to be downloaded by nodes $b$, $c$ and $s$. The syncronization process depends on average download rate ($R$) and average block validation Rate ($R_v$) and block size $b$ (assuming a constant block size).
The synchronization time required for these new nodes joining these networks are defined as follows

\begin{align*}
T_b &= n_b \times \Big(\frac{b}{R} + T_p\Big) + \frac{n_b  \times b}{R_v} \\
T_c &= n_c \times \Big(\frac{b}{R} + T_p\Big) + \frac{n_c \times b}{R_v} \\
T_s &= n_s \times \Big(\frac{b}{R} + T_p\Big) + \frac{n_s \times b}{R_v} + n_s \times T_{proofs} 
\end{align*}
The first term represents the downloading of the blocks from the existing node where as the second term denotes the validation of the blocks. In a \textit{securePrune} network, the nodes need to validate the NI-PoE proofs in time $T_{proofs}$  for each block verification, whereas the number of blocks is less compared to the other two networks.

The bootstrap node, after obtaining its final \textit{state} from the most recent \textit{snapshot} and \textit{tailchain} could acts as a full node to bootstrap the new joining nodes.

\section{Results and Discussion}
Table \ref{table:values} lists the values of the parameters used for generating the results in this section.  See   Table \ref{table:symbols} for a description.

\begin{table}[!t]
\renewcommand{\arraystretch}{1.3}
\caption{ \\ Parameter values used for simulations}
\centering
\begin{tabular}{|c|l|}
%
\hline
\textbf{Parameter}&\textbf{value}\\
    \hline
    $n$ & $1000$  \\ 
    \hline
    $n_p$  & $8$ \\ 
    \hline
    $\lambda$ & 1/600 blocks/sec \\
    \hline
    $T_p$& $30$ msec \\
		\hline
	$b$ & $0.25$ MB \\
		\hline
	$R$ & $10$ Mbps \\ 
		\hline
    $k$ &  500 \\
    \hline 
    $\Delta_s$ & 1000 \\
   \hline
   $R_v$ & 0.25 Mbps \\
   \hline
   $T_{proofs}$ & 0.35 sec \\
   \hline
    
\hline
\end{tabular}
\label{table:values}
\end{table}
We have conducted an event-driven simulation using python
by generating events as per information propagation protocol \cite{info} of bitcoin for propagating a block from miner to reach the entire network. The events are classified as \textit{inv} -  sending a new block hash invitation, \textit{getblock} - requesting a new block, \textit{block} - sending a block to its peers and \textit{addblock} -  adding a received block to its local copy of blockchain.

We have simulated for a duration of $70$ days (equivalent to $10000$) blocks with a block creation rate of $\lambda = \frac{1}{600}$ ($1$ block per every $10$ minutes) similar to bitcoin block generation rate. We have chosen $13$ nodes as miners with hash rates as per hash distribution shown in \cite{hashrate}.

\begin{figure}[t]
    \includegraphics[width=\columnwidth]{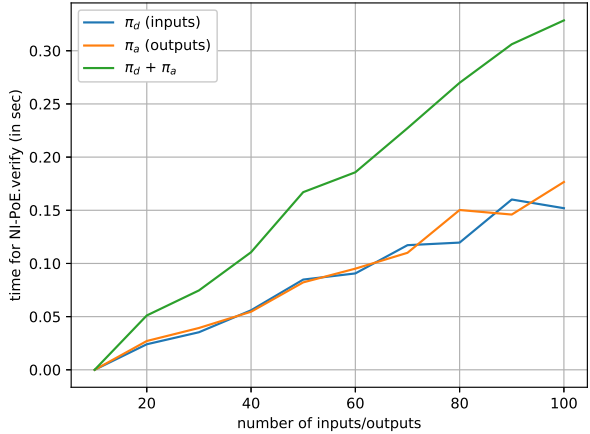}
  
  \caption{Verification time of NI-PoE 
  proofs Vs Number of \\ inputs/outputs transactions per block} 
\label{fig:proofs}
\end{figure}

Fig. \ref{fig:proofs} show the time required for a full node to verify NI-PoE proofs ($\pi_d$ and $\pi_a$) with  respect to the number of deleted sources of inputs ($|S_d|$) and number of added outputs ($|S_a|$). The verification time is very less ($\approx 0.35$ sec for $100$ inputs and $100$ outputs) compared to the block-creation time ($600$ sec).

\begin{figure}[t]
    \includegraphics[width=\columnwidth]{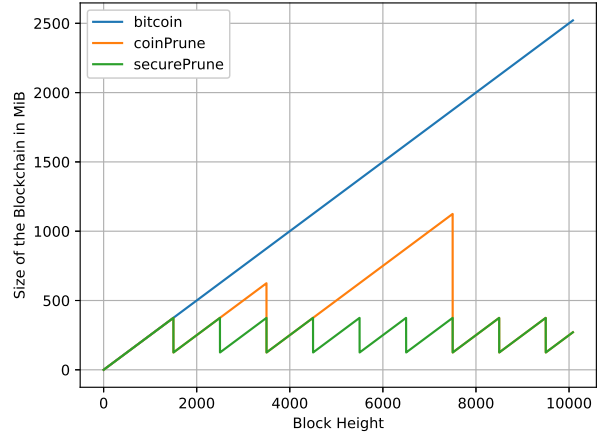}
  
  \caption{Storage comparisons of a nodes \\ in different protocols}
\label{fig:storage}
\end{figure}

Fig. \ref{fig:storage} show the total blockchain size of the nodes with respect to the block height. We have chosen the $1000$ blocks between the two consecutive \textit{snapshot}s and $500$ as number of confirmations (As the number of confirmations required for  double-spend to succeed by 
an attacker (with fraction of hash rate $q=0.45$) with a probability $< 10^{-4}$ is $462$ as per the calculations given in \cite{bitcoin}) for pruning the blocks prior to the \textit{snapshot}. The nodes prune old blocks at heights $1000+500c$ ($c = 1,2,3,\dots$).


\textit{Denial-of-Service attack on coinPrune protocol:} There is possible DOS attack on \textit{coinPrune} \cite{coinprune} by the miners in the network. Since the \textit{coinprune} requires a $k$ number  confirmations out of the number of blocks in \textit{reaffirmation window} to prune the blocks prior to the \textit{snapshot} . If the minimum requirement of $k$ confirmations not attained for any \textit{snapshot}, then the pruning could postponed to the next \textit{reaffirmation window}. We chose miners $m_{DoS} = \{1,7,8,10,12,13\}$ arbitrarily as DoS attackers with collective hash rate of $0.377$ ($\approx 38 \%$) and the other miners participating in the reaffirmations are $m_{reaffirm} = \{2,3,4,5,6,9,11\}$ have collective hash rate of $0.623$ ($\approx 62 \%$). We chose $300$ ($60 \%$ of size of reaffirmation window) confirmations out of \textit{reaffirmation window} of size $500$ blocks. The \textit{coinPrune} shows the pruning at block heights $3500$ and $7500$. In this case, the pruning of the \textit{coinPrune} node postponed a duration of $1000$, $2000$ blocks due to DoS attack by the above mentioned miners even though the number of confirmations ($300$) chosen are less than the number of confirmations ($500$) chosen for \textit{securePrune} block.

For values given in TABLE \ref{table:values},  the simulation results in Fig. \ref{fig:storage} show the maximum storage of \textit{securePrune} node is approaximately $400$ MiB $((\Delta_s+k) \times b$) for a block size of $0.25$ MiB, while the size of the bitcoin full node increases with block height. The reults show that $85\%$ reduction in the  the storage space of a \textit{securePrune} node compared to bitcoin nodes.

\begin{figure}[t]
    \includegraphics[width=\columnwidth]{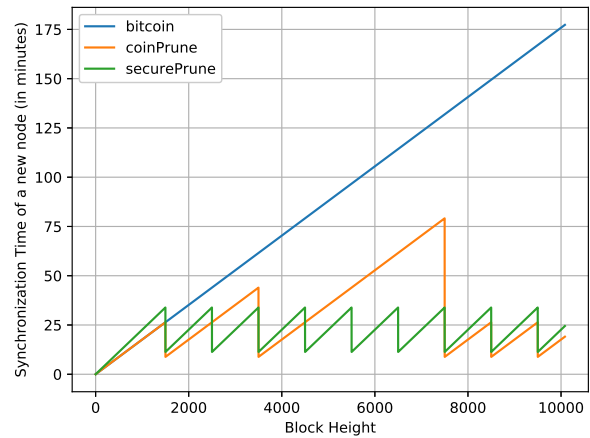}
  
  \caption{The bootstrapping time of new nodes  with respect to Block Height}
\label{fig:sync}
\end{figure}

Fig. \ref{fig:sync} show the time required for a bootstrapping node to synchronize with the existing nodes in the network. We hardcoded the block validation rate $R_v$ (depends on the processing speed of a node) and proofs verification time $T_{proofs}$ (from Fig. \ref{fig:proofs}) in the simulation. The synchronization time linearly proportional to the number of blocks present in the chain at the time of joining a new node. The  diiference in synchronization time of new nodes in \textit{coinPrune}  and \textit{securePrune} after pruning is due to the extra time ($T_{proofs}$) required for a new node in \textit{securePrune} to verify the NI-PoE proofs. Fig. \ref{fig:sync} show a significant reduction in synchronization time for a new node joining  \textit{securePrune} network compared to nodes joining the other two protocols.

Note: The results shown in Fig. \ref{fig:storage} and Fig. \ref{fig:sync} are obtained for different runs of simulations.

\section{Conclusion and Future Work}
In this paper, we show the periodic and secure pruning of the blocks prior to a certain block height based on the RSA accumulators. We proposed algorithms for generation of a block and validation of the block using NI-PoE proofs and accumulator state for securing the state of the blockchain along with transactions of the blocks. Through simulation results, we show the reduction in the storage space of a node in the proposed protocol which in turn reduce the synchronization time required to bootstrap a new node.

In future, we explore the exchanging of a  \textit{snapshot} from an existing node during the bootstrap of a new node while the state of the serving node changes with the creation of new blocks. We also consider the trade-off between the block generation, verification  with efficient NI-PoE proofs and number of transactions in the block. 

\bibliography{IEEEabrv,COMSNET2021.bib}

\end{document}